%% file: main.tex
\documentclass[sigconf,screen]{acmart}

\pdfoutput=1
\AtBeginDocument{%
  }

\usepackage{easyReview}
\usepackage{pifont}
\usepackage{tcolorbox}
\usepackage{tablefootnote}
\usepackage{enumitem}         
\usepackage{listings}
\usepackage{fancybox}
\usepackage{pifont}
\usepackage{multirow}
\usepackage{xcolor}
\usepackage{colortbl}
\usepackage{algorithmic}
\usepackage{algorithm}
\usepackage{setspace}
\usepackage{pgfplots}
\usepackage{pgf-pie}
\usepgfplotslibrary{statistics}
\usepackage{caption}
\usepackage{makecell}
\usepackage{arydshln}
\usepackage[normalem]{ulem}

\definecolor{mygray}{gray}{.9}
\definecolor{mycyan}{cmyk}{.3,0,0,0}
\definecolor{mauve}{rgb}{0.58,0,0.82}
\definecolor{dkgreen}{rgb}{0,0.6, 0}
\definecolor{gray}{rgb}{0.5,0.5,0.5}
\definecolor{lightgray}{rgb}{0.95,0.95,0.95}
\definecolor{ManiBlue}{HTML}{000080}
\definecolor{ManiRed}{HTML}{B22222}
\definecolor{ManiGray}{HTML}{9D9EA3}
\definecolor{ManiPurple}{HTML}{8074C8}
\definecolor{brown}{HTML}{bcaba4}
\definecolor{crashRed}{HTML}{d81e06}
\definecolor{easyRule}{HTML}{80c5a2}
\definecolor{hardRule}{HTML}{f27973}
\definecolor{deepGray}{HTML}{3F3F3F}

\setenumerate[1]{itemsep=0pt,partopsep=0pt,parsep=\parskip,topsep=5pt,leftmargin=15pt}
\setitemize[1]{itemsep=0pt,partopsep=0pt,parsep=\parskip,topsep=5pt,leftmargin=15pt}
\setdescription{itemsep=0pt,partopsep=0pt,parsep=\parskip,topsep=5pt,leftmargin=15pt}

\newcommand{\tabincell}[2]{\begin{tabular}{@{}#1@{}}#2\end{tabular}}
\newcommand{\ourtool}[1]{AutoChecker}

\newcommand*\circled[1]{\scalebox{0.8}{\tikz[baseline=(char.base)]{
\node[anchor=text, shape=circle,fill=deepGray, inner sep=0pt, minimum size=1.2em] (char) { \textbf{}\textcolor{white}{#1}};}}}
\newcommand*\roundedrect[1]{\scalebox{0.8}{\tikz[baseline=(char.base)]{
\node[anchor=text, shape=rectangle, fill=deepGray, inner sep=0pt, minimum width=2em, minimum height=1.2em, rounded corners=5pt] (char) {\textbf{}\textcolor{white}{#1}};}}}

\setcopyright{none} 
\renewcommand\footnotetextcopyrightpermission[1]{}

\begin{document}
\abovedisplayskip=6pt
\abovedisplayshortskip=6pt
\belowdisplayskip=6pt
\belowdisplayshortskip=6pt
\title{Write Your Own Code Checker: An Automated Test-Driven Checker Development Approach with LLMs}

\author{Jun Liu}
\authornote{Both authors contributed equally to this research.}
\email{liuj2022@ios.ac.cn}
\additionalaffiliation{%
 \institution{School of Advanced Interdisciplinary Science, UCAS}
 \city{Beijing}
 \country{China}
}

\author{Yuanyuan Xie}
\authornotemark[1]
\email{xieyy@ios.ac.cn}
\additionalaffiliation{%
 \institution{School of Intelligent Science and Technology, Hangzhou Institute for Advanced Study, University of Chinese Academy of Science (UCAS)}
 \city{Hangzhou}
 \country{China}
}
\affiliation{%
 \institution{Key Laboratory of System Software (Chinese Academy of Sciences) and State Key Laboratory of Computer Science, Institute of Software, Chinese Academy of Sciences}
 \city{Beijing}
 \country{China}
}
\author{Jiwei Yan}
\authornote{Corresponding author.}
\email{yanjiwei@otcaix.iscas.ac.cn}

\author{Jinhao Huang}
\email{me@jinhaohuang.com}
\affiliation{%
 \institution{Technology Center of Software Engineering, Institute of Software, Chinese Academy of Sciences}
 \city{Beijing}
 \country{China}
}
\author{Jun Yan}
\email{yanjun@ios.ac.cn}

\author{Jian Zhang}
\email{zj@ios.ac.cn}
\affiliation{%
 \institution{Key Laboratory of System Software (Chinese Academy of Sciences) and State Key Laboratory of Computer Science, Institute of Software, Chinese Academy of Sciences}
 \city{Beijing}
 \country{China}
}

\input{tex/0_abstract}
\maketitle

\input{tex/1_introduction}
\input{tex/2_bg_motication}
\input{tex/3_method}
\input{tex/4_evaluation}
\input{tex/5_threats}
\input{tex/6_related_work}
\input{tex/7_conclusion}

\clearpage

\bibliographystyle{ACM-Reference-Format}
\bibliography{main}

\end{document}

%% file: tex/0_abstract.tex
\begin{abstract}
With the rising demand for code quality assurance, developers are not only utilizing existing static code checkers but also seeking custom checkers to satisfy their specific needs. Nowadays, various code-checking frameworks provide extensive checker customization interfaces to meet this need. However, both the abstract checking logic and the complex API usage of large-scale checker frameworks make this task challenging. To this end, automated code checker generation is anticipated to ease the burden of checker development. 
In this paper, we propose AutoChecker, an innovative LLM-powered approach that can write code checkers automatically based on only a rule description and a test suite. 
To achieve comprehensive checking logic, AutoChecker incrementally updates the checker's logic by focusing on solving one selected case each time. To obtain precise API knowledge, during each iteration, it leverages fine-grained logic-guided API-context retrieval, where it first decomposes the checking logic into a series of sub-operations and then retrieves checker-related API-contexts for each sub-operation.
For evaluation, we apply AutoChecker, five baselines, and three ablation methods using multiple LLMs to generate checkers for 20 randomly selected PMD rules. Experimental results show that AutoChecker significantly outperforms others across all effectiveness metrics, with an average test pass rate of 82.28\%. Additionally, the checkers generated by AutoChecker can be successfully applied to real-world projects, matching the performance of official checkers.
\end{abstract}

%% file: tex/1_introduction.tex
\section{Introduction}

Static code-checking tools play a crucial role in ensuring code quality by generating security reports based on a set of predefined rules. In practice, users often need to customize checkers to meet specific requirements~\cite{distefano2019scaling}. Recent studies~\cite{johnson2013don,tymchuk2018jit,mendoncca2022empirical} also emphasize the importance of tailoring code-checking tools to specific contexts, such as individual projects and security scenarios. For example, a survey of experienced developers~\cite{tymchuk2018jit} found that up to one-third of participants highlighted the need for project-specific rules. Thus, customizing static code checkers is important for quality assurance.

To meet this demand, many static analysis tools support custom checkers. For instance, PMD~\cite{PMD} and SonarQube~\cite{Sonarqube} allow users to write custom checkers in Java, while CodeQL~\cite{codeql} and other DSL-based tools~\cite{DBLP:journals/corr/abs-2401-01571} support custom queries in DSL formats. However, an empirical study~\cite{christakis2016developers} reveals that only 8\% of developers actually write custom checkers in practice. This gap stems from several obstacles of the task: the high complexity of checking frameworks~\cite{DBLP:conf/asplos/BrownNE16} (e.g., \textit{PMD's framework alone exceeds 30 KLOC}), massive framework-specific API knowledge, incomplete or unclear API documentation, and the non-trivial checking logic. These barriers make checker customization time-consuming and difficult, especially for users with urgent needs but limited tool familiarity.

Recently, the booming of Large Language Models (LLMs) has significantly advanced automatic code generation ~\cite{Shuai2022ReACC,wang2023review, uchitel2024scoping}. Inspired by this, we explore leveraging LLMs to auto-generate checker code, aiming to alleviate the burden on developers in writing custom checkers. 
Notably, several recent studies have combined LLMs with static checking tools for security issue detection. Specifically, some studies~\cite{wang2023boosting,li2024llm} leverage LLMs to infer source-sink specifications for specific projects and CWEs, while others~\cite{li2023assisting,chen2024utilizing,li2024enhancing} use LLMs to filter false positives reported by static checkers. However, these works focus on enhancing existing checkers rather than creating new ones for specific requirements. As far as we know, we are the first to automate custom checker development using LLMs.

\begin{figure}[!ht]
\setlength{\abovecaptionskip}{5pt}
\setlength{\belowcaptionskip}{-5pt}
\centering
\includegraphics[width=0.38\textwidth]{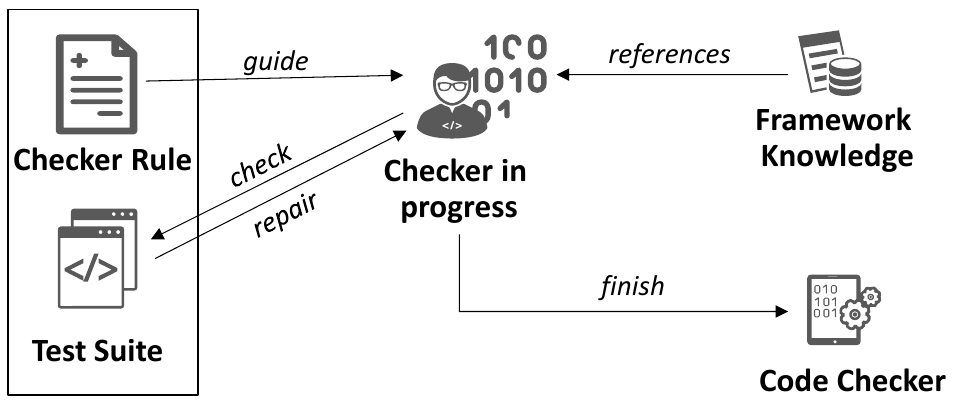}
\caption{Pipeline of the Manual Checker Development}
\label{fig:manualChecker_overview}
\end{figure}

Checker generation is more challenging and distinct from typical code-generation tasks. Fig.~\ref{fig:manualChecker_overview} illustrates the manual checker development process. When a custom checker is needed, the project manager provides an overall rule description for the rough goal and an adequate test suite to clarify the rule's detailed requirements. Developers then interpret the rule description and test suite to derive the correct checking logic and implement it using framework APIs based on their knowledge. Unlike typical code-generation tasks, where the target code (e.g., algorithm implementations or function code) is usually well-defined~\cite{wang2023review,liu2023your}, the complexity of expected checking logic and the extensive framework APIs make automated checker generation significantly more difficult. Specifically, we have to cope with the following two main challenges:

\begin{itemize}
    \item[\textbf{\textit{C1:}}] \textbf{Generating comprehensive checking logic covering diverse scenarios.} When leveraging LLMs to generate the comprehensive checking logic, both the rule description (for the overall goal) and the test suite (covering diverse scenarios) should be included as input. However, as the number of checking scenarios increases, the input information becomes excessive. This not only overwhelms the LLM’s ability to summarize the thorough logic across all scenarios but may also exceed the LLM’s token limit. Therefore, the comprehensive checking logic is hard to generate at once.
    \item[\textbf{\textit{C2:}}] \textbf{Retrieving precise API knowledge from high-level rule descriptions.} Developing code checkers requires a deep understanding of the framework's APIs. However, with thousands of APIs, identifying the precise ones for a specific checker is challenging. A common approach is to retrieve relevant APIs based on the rule description. However, this often fails due to the \textit{granularity mismatch} between high-level rule descriptions and specific API functionalities. This discrepancy makes precise API retrieval difficult, as also shown by the results of the Retrieval Augmented Generation (RAG) baseline in Section~\ref{RQ1}.
\end{itemize}

To address above challenges, we propose \textbf{\ourtool{}}, a novel approach to automatically generate static checkers from rule descriptions and test suites. First, to cover diverse scenarios, we mimic the manual checker development process (Fig.~\ref{fig:manualChecker_overview}), where developers iteratively validate and refine the checker against a test suite. We introduce the Test-Driven Checker Development (TDCD) approach, enabling \ourtool{} to refine the checker case by case, incrementally building comprehensive checking logic that fully aligns with the test (\textbf{\textit{C1}}). Second, to address the difficulty of retrieving precise API knowledge, \ourtool{} employs \textbf{Logic-guided API-context Retrieval} to extract checker-related API knowledge (\textbf{\textit{C2}}). Unlike common RAG approaches, which typically use rule descriptions as queries but often struggle with granularity mismatches, \ourtool{} decomposes the checking logic into discrete sub-operations and retrieves corresponding API contexts from two specialized databases: \textit{Meta-API DB} (semi-automatically constructed) and \textit{Full-API DB} (automatically constructed). This fine-grained retrieval ensures that precise API knowledge is extracted for each sub-operation, enabling accurate checker generation.

In this paper, we implement \ourtool{} on PMD~\cite{PMD}, a widely-used static analysis tool\footnote{Notably, \ourtool{} can be readily adapted to other AST-based tools that support custom checkers with minimal human effort, which is further discussed in Section~\ref{sec:threats}.}. To evaluate \ourtool{}, we randomly select 20 PMD built-in rules (10 easy and 10 hard). Experimental results show that our approach outperforms baselines across all metrics. Specifically, \ourtool{}-generated checkers achieve an average test pass rate of 82.28\% (84.70\% for easy rules and 79.86\% for hard ones), which is $2.93\times$ and $2.11\times$ higher than the simplest baseline NoCaseLLM and the best baseline $\mathrm{NoCaseLLM}^{RC}$, respectively. Also, we further evaluate practicality by applying \ourtool{}-generated checkers (that pass all tests) to five large-scale Java projects. The results show that \ourtool{} can write checkers performing equivalently to official ones when sufficient test cases are provided. We conclude our main contributions as follows:
\begin{itemize}
    \item We propose an automated test-driven checker development approach (TDCD), which uses an iterative generation pipeline to cope with the complex checking logic case by case.
    \item We develop a logic-guided API-context retrieval strategy and design a general Meta-Op set for fine-grained and precise API retrieval, which contains 354 atomic checking operations.
    \item We implement our approach into \ourtool{}, which can automatically develop custom code checkers based on the given rule and test suite. The experimental results show that checkers generated by \ourtool{} 
    greatly outperform baseline methods across all effectiveness metrics. Comparable to the official checkers, they also achieve expected results on real-world, large-scale projects.
\end{itemize}

Both the code and the dataset of \ourtool{} are available at \textcolor{blue}{\url{https://github.com/SQUARE-RG/AutoChecker}}. To demonstrate intermediate LLM-generated checkers and results in each step of the checker-development cycle, we also provide a replay website for visualization at \textcolor{blue}{\url{https://autochecker.maskeduser.party}}.

%% file: tex/2_bg_motication.tex
\section{Background and Motivation}

In this section, we first briefly introduce the background of custom static code checkers, with a focus on the specific type (AST-based checkers) targeted in this paper. Then, we illustrate the challenges and our proposed solutions through a motivating example.

\subsection{Custom Static Code Checker}

Static code checkers are designed in static analysis tools to analyze code without executing it~\cite{stamelos2002code, chess2004static, ayewah2008using}. Many existing tools, such as PMD~\cite{PMD}, SonarQube~\cite{Sonarqube}, and CodeQL~\cite{codeql}, support the customization of code checkers. These custom checkers can be broadly categorized into two groups based on their analysis techniques: \textbf{AST-based} (by traversing Abstract Syntax Tree~\cite{binkley2007source}), and \textbf{flow-based} (by analyzing control- and data-flow). 
Flow-based checkers are heavyweight, so their customization typically involves enhancing specifications on predefined checkers~\cite{wang2023boosting,li2024llm}.
Compared to them, AST-based checkers are more lightweight with a straightforward checking process: traverse the AST of the target code, apply checking rules to relevant AST nodes, and report potential issues when a match is found.
So, they are easier to customize. To meet new customization demands, experienced developers can write new AST-based checkers from scratch (as shown in Fig.~\ref{fig:manualChecker_overview}).
These advantages also make AST-based checkers a preferred choice for software companies in quality assurance. Therefore, this paper specifically focuses on automating the development of AST-based checkers.

\subsection{Motivating Example}

\begin{figure*}[t]
\setlength{\abovecaptionskip}{5pt}
\setlength{\belowcaptionskip}{-5pt}
\centering
\includegraphics[width=0.88\textwidth]{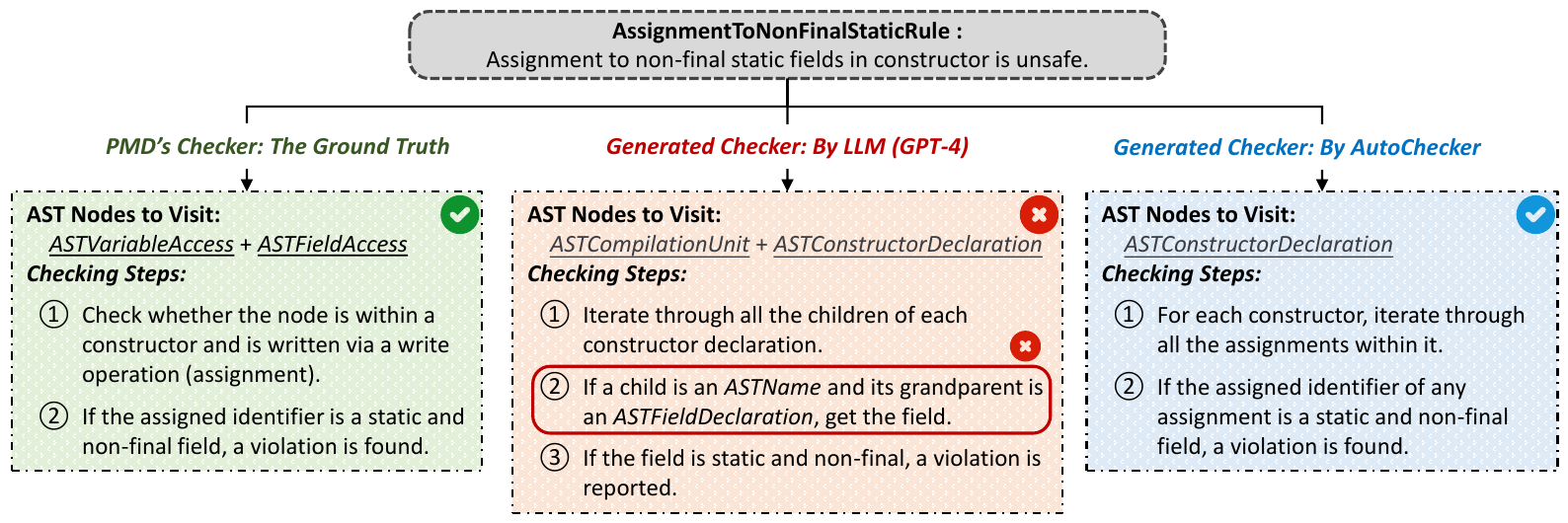}
\caption{A motivating example showing the concrete steps of the ground truth and auto-generated checkers for \textit{AssignmentToNonFinalStaticRule}. Specifically, the logic of the checker generated directly by LLM is incomplete.}
\label{fig:exp_ops}
\end{figure*}

PMD~\cite{PMD} is a popular AST-based static checking tool supporting 18 programming languages (primarily Java and Apex) with over 400 built-in rules. We use a PMD Java rule, \textit{AssignmentToNonFinalStaticRule}, as a motivating example. Its description states: \textit{``Assignment to non-final static fields in constructors is unsafe.''} The corresponding checker should report all unsafe assignments described by the rule.

First, we prompt multiple LLMs (GPT-4~\cite{openai_web}, DeepSeek-V3~\cite{liu2024deepseek}, etc.) to generate checkers for this rule by providing its description and full test suite. However, all generated checkers fail due to incomplete logic and compilation errors caused by hallucinated APIs. This highlights two key challenges in automated checker generation: (1) generating comprehensive checking logic (at the \textbf{Abstract Level}), and (2) invoking correct framework APIs (at the \textbf{Implementation Level}). Below, we detail the results from GPT-4.

At the \textbf{Abstract Level}, we compare the checking procedures in the LLM-generated checker and the ground truth. As shown in Fig.~\ref{fig:exp_ops}, the ground truth checker locates variable and field accesses within constructors and verifies if the referenced symbols are static and non-final. In contrast, the LLM-generated checker identifies unsafe fields starting from constructor declarations but only checks fields in \texttt{ASTFieldDeclaration}, missing unsafe fields in re-assignment expressions, resulting in incomplete logic. Despite providing sufficient test cases, the LLM struggles to generate comprehensive logic due to information overload from presenting many test cases at once. To address this, \ourtool{} introduces test-driven checker development, refining the checker's logic case by case. As shown in Fig.~\ref{fig:exp_ops}, \ourtool{} resolves the soundness issue by examining all assignment expressions within constructors, producing correct checking logic from a unique perspective compared to the ground truth.

\begin{figure}[!b]
\centering
\includegraphics[width=\linewidth]{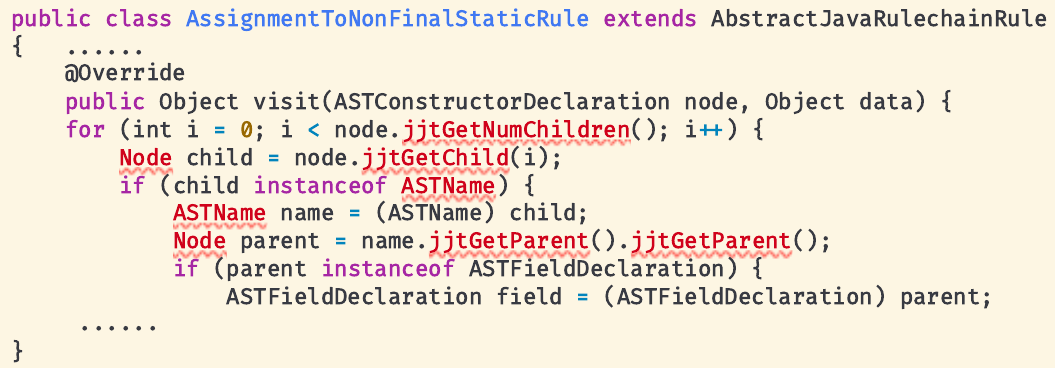}
\caption{A snippet of the LLM-generated checker for \textit{AssignmentToNonFinalStaticRule}, using the rule description and test suite as input, includes multiple hallucinated APIs.}
\label{fig:exp_pair}
\end{figure}

At the \textbf{Implementation Level}, we analyze the LLM-generated checkers' code. As shown in Fig.~\ref{fig:exp_pair}, when directly prompted to write a checker, the LLM often guesses framework APIs, leading to hallucinations like undefined method \texttt{jjtGetNumChildren} and class \texttt{ASTName}. Specifically, \textbf{41.7\% (5 out of 12)} of the APIs used are hallucinated, causing compilation errors. To address this, we then follow the common RAG pipeline~\cite{lewis2020retrieval}, retrieving framework APIs using the rule description as a query. However, due to the granularity mismatch between the high-level rule description and specific API functionality, \textbf{29.4\% (5 out of 17)} of the APIs remain hallucinated. Finally, by introducing fine-grained logic-guided API retrieval, \ourtool{} successfully generates a correct checker with \textbf{26} valid APIs, compiling and passing all tests. Notably, as API knowledge is provided, the number of APIs in the generated checker increases, as guessed APIs (often higher-level abstractions) are replaced with multiple concrete valid APIs.

%% file: tex/3_method.tex
\section{Methodology}
\label{sec:method}

\begin{figure*}[!t]
\setlength{\abovecaptionskip}{5pt}
\setlength{\belowcaptionskip}{-5pt}
\centering
\includegraphics[width=0.9\textwidth]{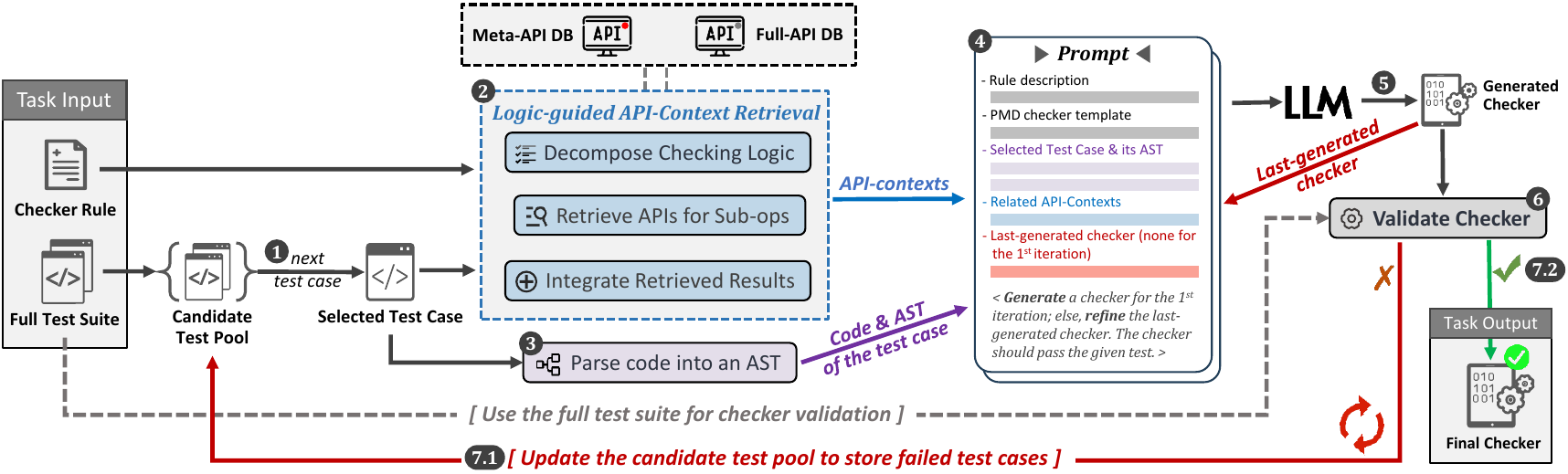}
\caption{Overview of the LLM-powered Test-Driven Checker Development in \ourtool{}}
\label{fig:LLMChecker_overview}
\end{figure*}

This section presents the detailed methodology of our proposed \ourtool{}. After showing the overall pipeline in Section~\ref{sec:overview}, Section~\ref{sec:retrieval} and Section~\ref{sec:cg} introduce the API-context retrieval and checker development approaches in detail.

\subsection{Overview}
\label{sec:overview}
Given a checker rule and its full test suite, \ourtool{} is designed to automatically generate the correct static checker following the \textbf{Test-Driven Checker Development (TDCD)} process. The overall pipeline of TDCD is shown in Fig.~\ref{fig:LLMChecker_overview}, in which \ourtool{} generates and refines the checker case by case. 

To start with, \ourtool{} maintains a candidate test pool to store test cases that have not yet been verified or passed. During each round of TDCD, a single test case is selected from this pool~\circled{1}. Using the selected test case and given checker rule, \ourtool{} employs the \textbf{Logic-guided API-Context Retrieval} approach to collect relevant API-contexts~\circled{2}. To ensure precision, \ourtool{} breaks down the checking logic into fine-grained sub-operations using LLM and retrieves the corresponding APIs respectively. Additionally, to obtain the accurate AST-based information of the test case, \ourtool{} utilizes a parser to get its AST~\circled{3}.

After preparing all the necessary input information, \ourtool{} constructs the checker-generation prompt \circled{4}, which consists of the \textit{rule description}, \textit{PMD checker template}, \textit{selected test case (both source code and AST)}, \textit{related API-contexts} and \textit{last-generated checker (not for the first round)}. By passing on the prompt to the LLM, a checker is generated for this round \circled{5}. To verify whether the generated checker is correct, it will then be validated with the full test suite \circled{6}. If the checker fails to pass all tests, \ourtool{} will update the candidate test pool to keep all the failed test cases and start the next iteration \roundedrect{7.1}. Otherwise, once the generated checker passes all tests or reaches a test-passing bottleneck, AutoChecker will terminate the TDCD process and output the final checker \roundedrect{7.2}.

\subsection{Logic-guided API-Context Retrieval}
\label{sec:retrieval}

As shown in Fig.~\ref{fig:LLMChecker_overview}, API-context Retrieval serves as a crucial module within the TDCD process, which is designed to provide accurate and sufficient API knowledge for checker generation. Inspired by Chain-of-Thought~\cite{wei2022chain,li2023structured} and Compositional API Recommendation~\cite{ma2024compositional}, we propose a fine-grained Logic-guided API-Context Retrieval approach. Specifically, \ourtool{} first uses the LLM to decompose the checker rule into a checking skeleton with sub-operations. Then, each sub-operation is used for individual API-context (API signatures and usages) retrieval and finally makes up the whole API-contexts. In this section, we sequentially explain the Logic-guided API-Context Retrieval approach in three parts: API Collection, Database Construction, and the Retrieval Process.

\subsubsection{Framework API Collection}

In general, framework APIs for AST-based checkers fall into the following three categories:
\begin{itemize}
    \item \textbf{Node-related APIs} perform concrete operations for specific AST nodes, e.g., obtaining the name of a method, etc.
    \item \textbf{Edge-related APIs} deal with connections and transitions between nodes, e.g.,  finding the closest parent AST node, etc.
    \item \textbf{Util-related APIs} offer utility functions that can be invoked anywhere, e.g., checking whether a type is abstract, etc.
\end{itemize}

In PMD, framework APIs\footnote{In the following text, we illustrate using PMD's Java code-checking APIs.} are defined in AST Node Classes (e.g., \texttt{ASTMethodDeclaration}) and Utility Classes (e.g., \texttt{JavaAstUtils}). Thus, we identify node- and edge-related APIs from AST Node Classes, while Util-related ones are collected from Utility Classes.

\ding{43} \textbf{Collecting Node-related and Edge-related APIs from AST Node Classes}. First, we map each AST Node Class (ANC) to its available APIs, including methods declared within the class and those inherited from its superclasses. Among all APIs, edge-related APIs, which handle general node-traversal functions, are primarily defined in the abstract ANC, \texttt{JavaNode}. From the available APIs of \texttt{JavaNode}, we identify edge-related APIs as those whose return value is another node. After filtering out these edge-related APIs, the remaining 
ones are categorized as node-related APIs.

\ding{43} \textbf{Collecting Util-related APIs from Utility Classes}. Each util-related API is a static method within a utility class characterized by a final modifier and a private constructor. By searching all the utility classes, we collect the util-related APIs.

Overall, the number of collected framework APIs in each type is shown in Tab.~\ref{tab:API-ctxs}. The significant number of APIs (over 11k) also underscores the necessity of precise retrieval. 

\begin{table}[!ht]
    \setlength{\abovecaptionskip}{5pt}
    \caption{PMD's Framework APIs of Each Type}
    \label{tab:API-ctxs}
    \centering
    \footnotesize
    \begin{tabular}{llc}
        \toprule
        API Type & Collect From & Number\\
        \midrule
        Node-related APIs & Concrete ANCs & 11,243\\ 
        Edge-related APIs  & Abstract ANC  & 21 \\ 
        Util-related APIs & Utility Classes & 377 \\ 
        \bottomrule
    \end{tabular}
\end{table}

\subsubsection{API-Context Database Construction}
\label{sec:collectapi}
Based on the collected framework APIs, we construct two API-context databases: \textbf{Full-API DB} and \textbf{Meta-API DB}. An API-context is defined as either an API's signature or usage snippet, paired with descriptive text (retrieval is based on semantic search of the text). The Meta-API DB is built using a crafted Meta-Op Set derived from the Full-API DB. We explain the process in three steps: \textit{Full-API DB Construction}, \textit{Meta-Op Set Preparation}, and \textit{Meta-API DB Construction}.

\ding{43} \textbf{Full-API DB Construction}. The Full-API DB is constructed using all three types of APIs. To generate the descriptive text for each API, we leverage the semantic information embedded in its signature. As demonstrated in Tab.~\ref{tab:API-des}, each descriptive text consists of three parts: the prefix, basic phrase, and comments.

\begin{table}[!b]
\setlength{\abovecaptionskip}{5pt}
\setlength{\belowcaptionskip}{-5pt}
\caption{Descriptive Text Generation for All Types of APIs}
\label{tab:API-des}
\centering
\setlength{\tabcolsep}{3pt}
\footnotesize
\begin{tabular}{lcr}
    \toprule
    \textbf{API  Type} & \textbf{Return Type}& \textbf{Descriptive Text (\textit{prefix+basic phrase+comments})}\\
    \midrule
    Node, Edge & \texttt{Boolean}      &  Check whether [className]$^s$ [methodName]$^s$ //cmt.\\
    Util &      \texttt{Boolean}      &  Check whether [methodName]$^s$ //cmt.\\ 
    Node, Edge &  \texttt{non-Boolean} &   [methodName]$^s$ of [className]$^s$ //cmt.\\ 
    Util &      \texttt{non-Boolean}  &   [methodName]$^s$ //cmt.\\ 
    \bottomrule
\end{tabular}
\footnotesize{\\$^s$ denotes splitting the name into individual words according to the CamelCase rule.\\ cmt. denotes the comments of each API for simplicity.}
\end{table}

First, we determine the \textbf{\textit{prefix}} of the descriptive text based on the API's return type. For an API with a \texttt{Boolean} return type, used for judgment, we add ``\textit{check whether}'' as the prefix of the descriptive text. For an API with a \texttt{non-Boolean} return type (e.g., \texttt{String}), used for data acquisition, the method name usually starts with an action word like ``\textit{get}'', so no additional prefix is required.

Then, we generate the \textbf{\textit{basic phrase}} based on the API's class and method names. Specifically, we split names into individual words based on the \textit{CamelCase} naming rule and remove unnecessary or repetitive terms (e.g., AST). For example, the class \texttt{ASTStringLiteral} yields the basic phrase ``\textit{String Literal}'', while the method \texttt{isEmpty} produces ``\textit{is empty}''. Notably, for util-related APIs, class names (e.g., \texttt{JavaAstUtil}) are typically omitted, as they often lack relevance to the API's concrete functionality. 

To enhance the descriptive text, we also extract \textbf{\textit{comments}} (docs) of the APIs and append them to the end of the description text, prefixed with ``//''. Irrelevant comments, such as those related to exceptional conditions or authorship, are filtered out. 

Finally, the prefix, basic phrase, and comments are combined to form the descriptive text of each API. Based on that, we construct the Full-API DB, where each element is a \textbf{description-signature} pair with the descriptive text and signature of an API. Fig.\ref{fig:fullapiexp} gives an example element for \texttt{isEmpty} in the Full-API DB.

\begin{figure}[!h]
\centering
\setlength{\abovecaptionskip}{5pt}
\setlength{\belowcaptionskip}{-5pt}
\begin{lstlisting}[numbers=none, basicstyle=\footnotesize, columns=fullflexible, escapeinside={||}]
|\textbf{\textcolor{ManiBlue}{\ding{252} Description-Signature Pair:}}| 
Description (|\textit{descriptive text}|): "Check whether string literal is empty"
API-context (|\textit{API signature}|): 
"net.sourceforge.pmd.lang.java.ast.ASTStringLiteral: 
public java.lang.Boolean isEmpty() //True if the constant value is empty."
\end{lstlisting}
\caption{An Example Element in Full-API DB}
\label{fig:fullapiexp}
\end{figure}

When using the Full-API DB for retrieval, we focus retrieval efforts on node- and util-related APIs and directly include all the edge-related API-contexts to the retrieved result. Edge-related APIs, which provide AST-traversing functions, are usually limited in number (21 for PMD as shown in Tab.~\ref{tab:API-ctxs}) but fundamental. Thus, we treat them as essential information to be provided by default.

\ding{43} \textbf{Meta-Op Set Preparation}. For real-world scenarios, framework APIs vary widely in encapsulation granularity, both within and across frameworks. This inconsistency makes it hard to reliably find the correct APIs solely based on the Full-API DB, which may lead to mismatches or overlaps. Thus, we need a more standardized API-context database (Meta-API DB). To meet this, we propose an abstraction layer, the Meta-Operation Set (Meta-Op Set), designed to unify API-context granularity across frameworks.

Specifically, the Meta-Op Set contains meta-operations (meta-ops) with basic functionalities commonly used for code-checking tasks. To get a comprehensive Meta-Op Set, we invited three developers with more than two years of checker-development experience for the collection. The first developer collected and organized most meta-ops into categories according to their experience across various checking frameworks (mainly based on PMD and CodeQL), and the other two brainstormed to supplement them. As shown in Fig.~\ref{fig:meta-ops}, the Meta-Op Set contains \textbf{354} meta-ops in \textbf{14} categories. We have open-sourced the Meta-Op Set in our project repository. 

\begin{figure}[!hb]
\setlength{\abovecaptionskip}{5pt}
    \centering
    \begin{tikzpicture}[scale=0.17]
        \tikzset{
             lines/.style={draw=gray!80, thin},
        }
        \pie[
            text=pin,
            sum=auto,
            radius=5.5, 
            font=\footnotesize,
            every only number node/.style={text=black},
            style={lines},
            color={red!30, blue!30, green!30, yellow!30, cyan!30, magenta!30, orange!30, violet!30, gray!30, lime!30, teal!30, pink!30, brown!30, purple!30}
        ]
        {
            27/Java Feature,
            36/Class,
            77/Method,
            48/Method Call,
            14/Control Stmt,
            33/Field,
            26/Local Var,
            30/Var Usage,
            9/Exception,
            15/Array,
            2/Object,
            21/Expression,
            15/Literal,
            1/Multi-thread
        }
    \end{tikzpicture}
    \captionof{figure}{Category of Operations in the Meta-Op Set}
    \label{fig:meta-ops}
\end{figure}

\ding{43} \textbf{Meta-API DB Construction}. Using the Meta-Op Set as a foundation, we construct the Meta-API Database (Meta-API DB), where each entry pairs a meta-operation (meta-op) with its corresponding API-context (either API signature or usage snippet). 

For each meta-op, we first search the Full-API DB to identify API descriptions that semantically align with the meta-op’s functionality. Once a match is found, we extract the associated API signature as the API-context for that meta-op. Otherwise, if no API descriptions match the given meta-op, we manually craft an implementation code snippet to fulfill the meta-op's functionality as its API-context. Overall, the API-contexts in Meta-API DB are in the form of \textbf{operation-signature} pairs and \textbf{operation-snippet} pairs. We provide two examples in Fig.~\ref{fig:metadbexp}.

\begin{figure}[!hpt]
\centering
\setlength{\abovecaptionskip}{5pt}
\setlength{\belowcaptionskip}{-10pt}
\begin{lstlisting}[numbers=none, basicstyle=\footnotesize, columns=fullflexible, escapeinside={||}]
|\textbf{\textcolor{ManiBlue}{\ding{252} Operation-Signature Pair:}}| 
Meta-op: "Get the name of class"; Category: "Class".
API-context (|\textit{API signature}|): 
"net.sourceforge.pmd.lang.java.ast.ASTClassOrInterfaceDeclaration: 
public java.lang.String getSimpleName()"

|\textbf{\textcolor{ManiBlue}{\ding{252} Operation-Snippet Pair:}}| 
Meta-op: "Check whether the return type of method is int"; Category: "Method".
API-context (|\textit{code snippet}|):
"import net.sourceforge.pmd.lang.java.ast.ASTMethodDeclaration;
import net.sourceforge.pmd.lang.java.types.JPrimitiveType;
public boolean isReturnValueIntType(ASTMethodDeclaration m) {
  return m.getResultTypeNode().getTypeMirror()
          .isPrimitive(JPrimitiveType.PrimitiveTypeKind.INT);
}"
\end{lstlisting}
\caption{Example Elements in Meta-API DB}
\label{fig:metadbexp}
\end{figure}

\subsubsection{API-Context Retrieval Process}
\begin{figure}[!hbt]
\setlength{\abovecaptionskip}{5pt}
\setlength{\belowcaptionskip}{-5pt}
\centering
\includegraphics[width=\linewidth]{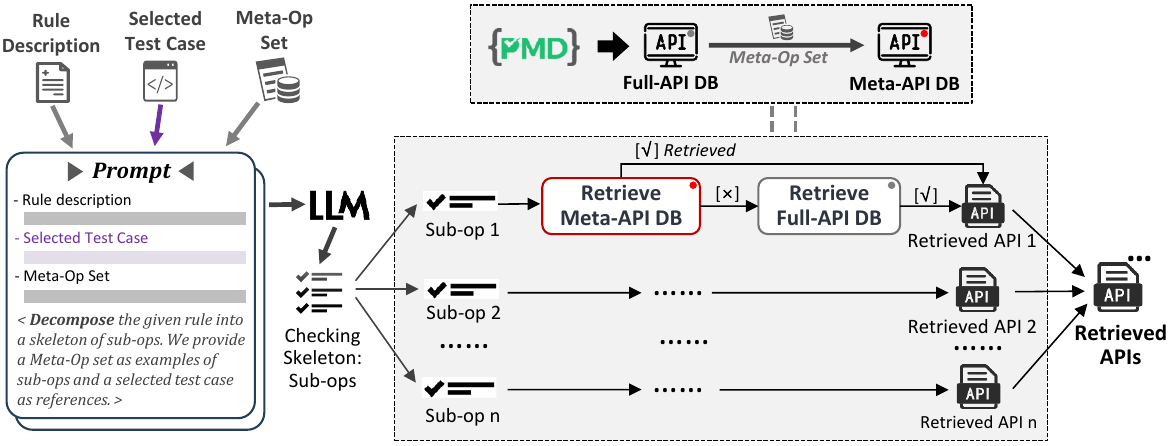}
\caption{Pipeline of the Logic-guided API-context Retrieval}
\label{fig:retrieval}
\end{figure}

With the constructed DBs, \ourtool{} retrieves related API-contexts based on the checker rule and a given test case. To start with, all 21 edge-related API-contexts from the Full-API-DB are directly added to collected API-contexts, as mentioned in Section~\ref{sec:collectapi}. Then, \ourtool{} leverages the Logic-guided API-context Retrieval approach to retrieve related node- and util-related API-contexts, which is shown in Fig.~\ref{fig:retrieval}. 

First, \ourtool{} generates a checking skeleton by decomposing the checker rule into sub-operations (sub-ops). Given the checker rule, test case and Meta-Op Set as inputs, \ourtool{} leverages the LLM to make the split. Specifically, the Meta-Op Set serves as references to sub-ops, which guides the LLM to generate sub-operations under the similar granularity of meta-ops. In Fig.~\ref{fig:retrieval}, the overall decomposition prompt is also demonstrated.

Next, \ourtool{} fetches API-contexts for each sub-op using both the Meta-API and Full-API DBs. During each retrieval process, the sub-op serves as the query to find the API-context with the highest semantic similarity score. If the score falls below a set threshold, the retrieval fails and returns \texttt{None}. \ourtool{} first queries the Meta-API DB. If unsuccessful, it then searches the Full-API DB. Note that, before querying the Full-API DB, \ourtool{} filters out irrelevant node-related APIs for higher precision and efficiency. Here, APIs defined in AST node classes that don't appear in the test case's AST are deemed irrelevant. Finally, all relevant API-contexts, both foundational and retrieved, are gathered.

\subsection{Test-Driven Checker Development}
\label{sec:cg}
In this part, we focus on the technical details of the TDCD process. Specifically, Algorithm~\ref{alg:tdcd} shows how to get the final checker iteratively, which follows the overall pipeline in Fig.~\ref{fig:LLMChecker_overview}.

\begin{algorithm}
	\renewcommand{\algorithmicrequire}{\textbf{Input:}}
	\renewcommand{\algorithmicensure}{\textbf{Output:}}
	\footnotesize
    \captionsetup{font=small}
	\caption{Algorithm of Test-Driven Checker Development (TDCD)}
	\label{alg:tdcd}
	\begin{algorithmic}[1]
		\REQUIRE $\sf r$: the checker rule description, $\sf T_a$: the full test suite with all tests
		\ENSURE $\sf c_f$: the final checker, $\sf pr_f$: the test pass rate for the final checker
		\STATE \emph{Load the checker template $\sf \overline{C}$}
		\STATE $\sf T_c \gets T_a, c \gets \texttt{None}$
		\textcolor{blue}{\COMMENT{initialize the candidate test pool $\sf T_c$ and checker $\sf c$}}
		\STATE $\sf T_p \gets \{\}, T_s \gets \{\}$
 		\textcolor{blue}{\COMMENT{record the passed tests in $\sf T_p$ and skipped tests in $\sf T_s$}}
		\WHILE{$\sf |T_c| > 0$}
		    \STATE $\sf t \gets \mathit{pickNextTest}(T_c)$ \label{alg:tdcd5}
		    \STATE $\sf K_{api} \gets \mathit{retrieveAPIContexts}(r, t)$ \label{alg:tdcd7}
		    \textcolor{blue}{\COMMENT{use logic-guided API-context retrieval}}
		    \STATE $\sf ast \gets \mathit{parseAST}(t)$ \label{alg:tdcd8}
		    \STATE $\sf j \gets 0$ \label{alg:tdcd9}
		    \textcolor{blue}{\COMMENT{the number of retries for $\sf t$}}
		    \WHILE{$\sf j < MAX\_RETRY\_TIMES$}
		        \IF{$\sf c = \mathrm{None}$}
            		\STATE $\sf c \gets \mathit{genInitialChecker}(r, t, ast, \overline{C}, K_{api})$ \label{alg:tdcd12}
            		\textcolor{blue}{\COMMENT{LLM-based generation}}
        		\ELSE
        		    \STATE $\sf c \gets \mathit{refineLastChecker}(r, t, ast, \overline{C}, K_{api}, c)$ \label{alg:tdcd14}
            		\textcolor{blue}{\COMMENT{LLM-based refinement}}
        		\ENDIF
        		\STATE $\sf rep \gets \mathit{validateChecker}(c, T_a)$ \label{alg:tdcd16}
		        \textcolor{blue}{\COMMENT{get the validation report}}
        		\IF{$\sf t \in rep.passedtests$ and $\sf rep.failedtests \cap T_p = \emptyset$}
            		\STATE  \textbf{break}
            		\textcolor{blue}{\COMMENT{the checker passes $\sf t$ without regression errors}}
            	\ENDIF
            	\STATE $\sf j \gets j+1$ \label{alg:tdcd20}
		    \ENDWHILE \label{alg:tdcd21}
    		\IF{$\sf j = MAX\_RETRY\_TIMES$} \label{alg:tdcd22}
    		\STATE $\sf T_s.\mathit{add}(t)$ \label{alg:tdcd23}
    		\textcolor{blue}{\COMMENT{skip $\sf t$ if it reaches the retry limit}}
    		\ENDIF
    		\STATE $\sf T_p \gets rep.passedtests, T_c \gets rep.failedtests \setminus T_s$ \label{alg:tdcd25}
    		\textcolor{blue}{\COMMENT{update test sets}}
		\ENDWHILE
        \STATE $\sf c_f \gets c, \sf pr_f \gets rep.pr$	
        \textcolor{blue}{\COMMENT{return the final checker and test pass rate}}
	\end{algorithmic}  
\end{algorithm}

\subsubsection{Prompt Settings}
In each round of TDCD, \ourtool{} writes a checker based on a selected test and the checker rule. There are two types of prompts in TDCD: one for initial checker generation and the other for iterative checker refinement.

\ding{43} \textbf{Prompt for Initial Generation}. In the $1^{st}$ round, the prompt instructs the LLM to generate a rule-specific checker capable of passing the provided test using the following input on line~\ref{alg:tdcd12}.

\begin{figure}[!ht]
\setlength{\abovecaptionskip}{5pt}
\setlength{\belowcaptionskip}{-5pt}
\centering
\includegraphics[width=0.9\linewidth]{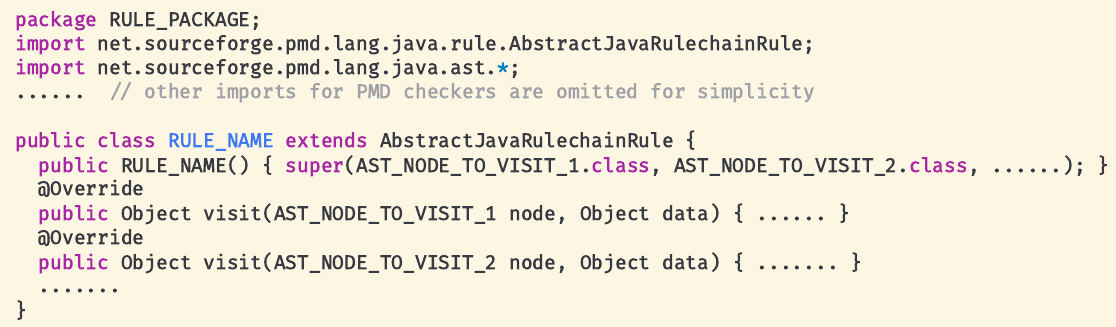}
\caption{Simplified PMD Checker Template}
\label{fig:pmd-tmp}
\end{figure}

\begin{itemize}[label={\ding{72}}]
    \item \textbf{\textit{Rule description}}. It is derived from the original input.
    \item \textbf{\textit{Test case code}}. It is picked from the candidate test pool (line~\ref{alg:tdcd5}).
    \item \textbf{\textit{Test case AST}}. Since AST information is crucial for AST-based checking, \ourtool{} extracts the test’s AST using PMD’s built-in parser (line~\ref{alg:tdcd8}). To clearly link AST nodes to their source code, \ourtool{} also retains the concrete names of AST nodes parsed from identifiers. For instance, the AST node \texttt{ASTClassDeclaration} parsed from the method name ``length'' is augmented as ``ASTMethodDeclaration(length)''.
    \item \textbf{\textit{Related API-contexts}}. \ourtool{} adopts the API-Context Retrieval to retrieve related API-contexts based on the checker rule and cleaned test case on line~\ref{alg:tdcd7}, introduced in Sec.~\ref{sec:retrieval}.
    \item \textbf{\textit{Checker template}}. We manually summarize a PMD checker template from existing checkers, which is shown in Fig.~\ref{fig:pmd-tmp}.
\end{itemize}

\ding{43} \textbf{Prompt for Iterative Refinement}. In subsequent rounds, the prompt is designed for checker refinement. It instructs the LLM to refine a given rule-specific checker to pass the selected test case. Compared to the initial generation prompt, this one also includes the \ding{72} \textbf{\textit{last-generated checker}} as input.

Notably, after generating the checker using the above prompts, \ourtool{} employs a simple strategy to prevent import errors. Specifically, it replaces the import section of the generated checker code with default imports, matching those in the template (Fig.~\ref{fig:pmd-tmp}). This ensures that all required packages are correctly imported.

\subsubsection{Checker Development Cycle}

The TDCD cycle follows an iterative refinement process. Throughout the cycle, \ourtool{} dynamically maintains three test sets as follows.
\begin{itemize}
    \item $\sf T_c$ is the candidate test pool with unprocessed and failed tests.
    \item $\sf T_p$ is a test set that records all passed tests.
    \item $\sf T_s$ is a test set that records all skipped tests. In a single round, sometimes the LLM may fail to generate a checker that passes the given test case within allowed attempts, \ourtool{} then skips this test to prevent blocking the cycle.
\end{itemize}

To start with, the cycle begins by initializing $\sf T_c$ with all tests from the full suite $\sf T_a$. Then, \ourtool{} selects a single test from $\sf T_c$ in each round of the cycle to guide the checker development process on lines~\ref{alg:tdcd5}-\ref{alg:tdcd20}. For each round, the generated checker will be validated with the full test suite on line~\ref{alg:tdcd16}. Note that \ourtool{} ensures that each newly generated checker in every iteration should pass the given test case without affecting the already passed test cases (without regression errors). If not, \ourtool{} will re-query the LLM to re-generate the checker within allowed retry attempts on lines~\ref{alg:tdcd9}-\ref{alg:tdcd21}. After validation, all test sets are updated on lines~\ref{alg:tdcd22}-\ref{alg:tdcd25}. Specifically, passed tests are moved to the $\sf T_p$, while persistently failing tests (after maximum attempts) are added to $\sf T_s$. 
Besides, $\sf T_c$ is updated with the failed tests, excluding skipped ones in $\sf T_s$.

Finally, the cycle terminates when $\sf T_c$ becomes empty, indicating all tests have been either validated or skipped. The final checker $\sf c_f$ and its test pass rate $\sf pr_f$ are derived from the last validation results.

%% file: tex/4_evaluation.tex
\section{Evaluation}

We conduct extensive experimental evaluations of \ourtool{} to address the following research questions:

\begin{itemize}
    \item \textbf{RQ1 (Effectiveness)}: Can \ourtool{} effectively generate high-quality code checkers?
    \item \textbf{RQ2 (Ablation Study)}: How do different strategies contribute to \ourtool{}’s effectiveness?
    \item \textbf{RQ3 (Cost)}: Can \ourtool{} develop checkers cost-effectively?
    \item\textbf{RQ4 (Practicality)}: How do \ourtool{}-generated checkers perform on real-world projects?
\end{itemize}

\subsection{Evaluation Setup}
\label{sec:configure}

\subsubsection{Implementation Settings}

In this paper, we build \ourtool{} specifically for \textit{PMD}, an open-source AST-based code-checking tool known for its effectiveness and ease of use~\cite{liu2023comprehensive}. Specifically, we used the latest version \texttt{7.0.0-rc4} when we started our work. 

\ourtool{} is implemented on \textit{LangChain}~\cite{Langchain_web}, a widely-used framework for LLM-based applications. For the API-context retrieval module, we use the SOTA open-source embedding model \textit{bge-large-en-v1.5}~\cite{bge_embedding} from BAAI~\cite{baai_web} and design two similarity score thresholds referring to our experience and previous work~\cite{zhang2023retrieve, liu2024fix}: 0.85 for Meta-API matching and 0.8 for API-context searching. In the checker development cycle, we set \texttt{MAX\_RETRY\_TIMES} as 5 for each round of checker generation. Currently, \ourtool{} supports two working modes: \textit{writing checkers from scratch} and \textit{incrementally}. In the incremental mode, developers can enhance existing checkers by providing additional test cases, which will continuously trigger the test-driven checker development (TDCD) process.

To evaluate the effectiveness of \ourtool{}, we use multiple popular LLMs, including self-hosted and official ones, as follows:
\begin{itemize}
    \item Self-hosted LLMs:  Llama3.1 (\texttt{Llama-3.1-8B-Instruct})~\cite{llama_web} and Qwen2.5-Coder (\texttt{Qwen2.5-Coder-32B-Instruct-AWQ})~\cite{hui2024qwen2}.
    \item Official LLMs: GPT-4 (\texttt{gpt-4-0613})~\cite{openai_web} and DeepSeek-V3~\cite{liu2024deepseek}.
\end{itemize}

\subsubsection{Benchmark RuleSet}
\label{sec:bench}
\begin{figure}[!ht]
\setlength{\abovecaptionskip}{5pt}
\setlength{\belowcaptionskip}{-5pt}
\centering
\includegraphics[width=0.48\textwidth]{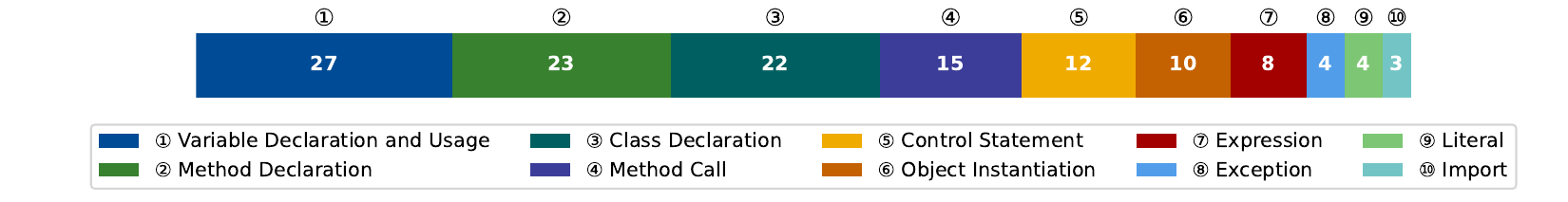}
\caption{Distribution of Classified PMD's Built-in Rules}
\label{fig:classify}
\end{figure}

The benchmark ruleset for evaluation is derived from the official built-in rules in PMD 7.0.0-rc's open-source repository~\cite{PMD_repo}. Initially, there are 132 built-in PMD Java rules. We exclude four rules that are either deprecated or undocumented\footnote{Excluded rules are \textit{ExcessiveMethodLength}, \textit{ExcessiveClassLength}, \textit{BeanMembersShouldSerialize}, and \textit{AbstractNamingConvention}.}. The remaining 128 rules are classified based on the primary ASTNode they check, as defined in their official implementations. Fig.~\ref{fig:classify} shows the distribution of rules across these reclassified categories.

For a clearer evaluation, we also divide the collected rules into \textbf{\textit{easy rules}} and \textbf{\textit{hard rules}} based on the implementation complexity of their official checkers. Statistically, we measure complexity by analyzing specific elements in the checker code: a rule is labeled as \textit{easy} if its checker's \textit{line count}, \textit{import statements}, \textit{method calls}, and \textit{control statements} are all below the average values across all built-in checkers, and if it uses fewer than one semantic class (from \texttt{pmd.lang.java.types} or \texttt{pmd.lang.java.symbols}). Rules not meeting these criteria are labeled as \textit{hard}.

Overall, we have 128 rules across 10 categories, evenly split into 64 easy and 64 hard rules. For evaluation, we randomly select 10 easy and 10 hard rules, ensuring each represents a unique category. Since PMD provides official test cases for each rule, we extract the default test suites for these 20 rules from PMD's website~\cite{PMD}. By default, these test cases are generally ordered by their difficulty, and we retain this order for \ourtool{}. Finally, the benchmark ruleset's details are summarized in Tab.~\ref{tab:benchmark}.

\begin{table}[!t]
\caption{Basic Information of the Benchmark RuleSet}
\label{tab:benchmark}
\centering
\scriptsize
\setlength{\tabcolsep}{1pt}
\renewcommand{\arraystretch}{1.2}
\begin{tabular}{>{\bfseries}c|l|c|l|c}
    \Xhline{0.8px}
      & \multicolumn{2}{c|}{\textbf{Easy Rules}} & \multicolumn{2}{c}{\textbf{Hard Rules}} \\
    \cline{2-5}
    \multirow{-2}{*}{\textbf{Category}} & \textbf{Rule Name}  & \textbf{\#TC} & \textbf{Rule Name} & \textbf{\#TC}  \\
    \hline
    Method Decl.        & \cellcolor{easyRule!15}\tabincell{l}{SignatureDeclareThrowsException}   & \cellcolor{easyRule!15}22      & \cellcolor{hardRule!15}MethodNamingConventions & \cellcolor{hardRule!15}12 \\
    \hline
    
    Method Call         & \cellcolor{easyRule!15}\tabincell{l}{InefficientEmptyStringCheck}   & \cellcolor{easyRule!15}18  & \cellcolor{hardRule!15}LiteralsFirstInComparisons  & \cellcolor{hardRule!15}33           \\
    \hline
    
    Class Decl.         & \cellcolor{easyRule!15}ExcessivePublicCount  & \cellcolor{easyRule!15}7  &  \cellcolor{hardRule!15}\tabincell{l}{ClassWithOnlyPrivateConstructors\\[-2.5pt]ShouldBeFinal}& \cellcolor{hardRule!15}22         \\
    \hline
    
    \tabincell{l}{Variable Decl.\\ and Usage}   & \cellcolor{easyRule!15}\tabincell{l}{UseStringBufferForStringAppends}  & \cellcolor{easyRule!15}28          & \cellcolor{hardRule!15}AssignmentToNonFinalStatic                    & \cellcolor{hardRule!15}6          \\
    \hline
    
    Exception       & \cellcolor{easyRule!15}ExceptionAsFlowControl  & \cellcolor{easyRule!15}7  & \cellcolor{hardRule!15}\tabincell{l}{AvoidThrowingNullPointerException}     & \cellcolor{hardRule!15}9          \\
    \hline
    
    Expression      & \cellcolor{easyRule!15}NullAssignment   & \cellcolor{easyRule!15}19   & \cellcolor{hardRule!15}BrokenNullCheck & \cellcolor{hardRule!15}25       \\
    \hline
    
    Control Stmt    & \cellcolor{easyRule!15}IdenticalCatchBranches   & \cellcolor{easyRule!15}7   & \cellcolor{hardRule!15}EmptyControlStatement  & \cellcolor{hardRule!15}31      \\
    \hline
    
    Object Inst.    & \cellcolor{easyRule!15}StringInstantiation & \cellcolor{easyRule!15}10   & \cellcolor{hardRule!15}\tabincell{l}{AvoidInstantiatingObjects\\[-2.5pt]InLoops}              & \cellcolor{hardRule!15}23       \\
    \hline
    
    Import          & \cellcolor{easyRule!15}ExcessiveImports   & \cellcolor{easyRule!15}2    & \cellcolor{hardRule!15}UnnecessaryImport & \cellcolor{hardRule!15}73     \\
    \hline
    
    Literal         & \cellcolor{easyRule!15}AvoidUsingOctalValues     & \cellcolor{easyRule!15}8       & \cellcolor{hardRule!15}AvoidDuplicateLiterals  & \cellcolor{hardRule!15}11      \\ 
    \Xhline{0.8px}
\end{tabular}
\footnotesize{\\\textbf{\#TC:} the number of test cases. \textbf{Abbr.}: Decl.$\rightarrow$Declaration, Inst.$\rightarrow$Instantiation}
\end{table}

\subsubsection{Baselines and Ablation Methods} 
\label{sec:baseline}
According to our knowledge, \ourtool{} is the first LLM-based approach for automated code checker generation, specifically for AST-based ones. Thus, we manually develop comprehensive baseline and ablation methods based on LLMs to demonstrate the effectiveness of \ourtool{}. 

For \textbf{RQ1}, we design five baselines to generate the checker at one time inspired by common practices in LLM-powered SE tasks~\cite{hou2024large}:
\begin{itemize}
\item \textbf{NoCaseLLM}: generates checkers using only the rule description and PMD's checker template, without test cases.
\item \textbf{AllCasesLLM}: generates checkers with the rule description, PMD checker template, and the full test suite. If the test suite exceeds the LLM's token limit, excess cases are dropped.
\item \textbf{$\rm \textbf{NoCaseLLM}^{\textbf{R}}$}: enhances NoCaseLLM with RAG, adding the top-k (default k=19, the mean API count of PMD's built-in checkers) APIs retrieved from the Full-API DB using the rule description as query.
\item \textbf{$\rm \textbf{NoCaseLLM}^{\textbf{C}}$}: enhances NoCaseLLM with Chain-of-Thought (CoT) prompting, the LLM is asked to ``\textit{first create a comprehensive checking skeleton and then generate the checker}''.
\item \textbf{$\rm \textbf{NoCaseLLM}^{\textbf{RC}}$}: enhances NoCaseLLM with both RAG and COT strategies.
\end{itemize}

For \textbf{RQ2}, we evaluate the impact of \ourtool{}'s two key strategies: the logic-guided API-context retrieval and the TDCD cycle (case-by-case iteration). We designed three ablation methods:
\begin{itemize}
\item \textbf{$\rm \textbf{AutoChecker}^{\textbf{WoI}}$}: removes the TDCD cycle, providing all test cases, their ASTs, and API-contexts at once. Excess tests are dropped, similar to AllCaseLLM.
\item \textbf{$\rm \textbf{AutoChecker}^{\textbf{WoR}}$}: removes the API-context retrieval but retains the TDCD cycle, prompting LLMs without API-contexts.
\item \textbf{$\rm \textbf{AutoChecker}^{\textbf{WoM}}$}: removes Meta-Op Set and Meta-API DB. For API-context retrieval, it splits logic into sub-ops based on the rule and test case and retrieves solely on Full-API DB.
\end{itemize}

In our evaluation, we run each method (including baselines and \ourtool{})  \textbf{three} times to account for LLM's randomness, and the best performance from each is collected for fair comparison.

\subsubsection{Metrics}
\label{sec:metric}
We design four types of metrics to evaluate a given approach in developing static code checkers.

\ding{117} $\textbf{Rule}_{pc}$: A rule is counted as $Rule_{pc}$ if the approach successfully generates a \underline{p}ass-\underline{c}ompilation checker for it. For the approach, the total number of such rules is recorded as \#${Rule}_{pc}$.

\ding{117} $\textbf{Rule}_{pot}$: A rule is counted as ${Rule}_{pot}$ if the approach generates a checker that \underline{p}asses at least \underline{o}ne of its \underline{t}est case. The total number of such rules is recorded as \#${Rule}_{pot}$.

\ding{117} $\textbf{Rule}_{pat}$: A rule is counted as ${Rule}_{pit}$ if the approach generates a checker that \underline{p}asses \underline{a}ll the \underline{t}est cases in its test suite. The total number of such rules is recorded as \#${Rule}_{pat}$.

\ding{117} $\textbf{TPR}$ and $\textbf{TPR}_{avg}$: For each rule, we record the test pass rate 
($\frac{number\ of\ passed\ test\ cases}{number\ of\ all\ test\ cases}$) of the generated final checker as $TPR$. $TPR_{avg}$ denotes the average pass rate across all rules.

\subsection{RQ1: Effectiveness Evaluation}
\label{RQ1}

Tab.~\ref{tab:RQ1_result} shows the main evaluation result of \ourtool{} and other baseline methods on the benchmark ruleset based on metrics defined in Section~\ref{sec:metric}. For each method, we record the result with the highest $\rm TPR_\textit{avg}$ across three runs for fair comparison. 

When paired with GPT-4, \ourtool{} outperforms all other baselines across different LLMs on all metrics. Specifically, it successfully generates checkers that can pass all tests for six rules, and at least one for all 20 rules (passing 278 test cases in total). Though the generated checkers cannot pass all tests for all the rules, they attain an 82.28\% average test pass rate ($\rm TPR_\textit{avg}$), indicating the method's remarkable effectiveness in generating usable checkers. 

In general, the performance of all methods (excluding ablation methods in this RQ) across various LLMs follows these rankings:
\small{
\begin{itemize}
    \item \textbf{\textit{LLM Rank}}: $\rm \ Llama3.1< \textrm{Qwen2.5-Coder} \lesssim \textrm{GPT-4} \lesssim \textrm{DeepSeek-V3}$
    \item \textbf{\textit{Method Rank}}: $\rm \ AllCasesLLM < NoCasesLLM < NoCaseLLM^C \\ < NoCaseLLM^R < NoCaseLLM^{RC} < \ourtool{}$.
\end{itemize}}

The LLM-rank result generally aligns with other LLM-evaluation studies~\cite{liu2023your,hui2024qwen2,liu2024deepseek}. The smallest model, Llama3.1, with limited code-related capability, often leads to compilation failures caused by syntax errors. In contrast, the other three LLMs, being more powerful, can generate test-passing checkers. Among them, DeepSeek-V3 excels in all baselines, while GPT-4 gets the best result for \ourtool{} (checkers generated with DeepSeek-V3 and GPT-4 pass the same number of tests but vary in test distribution over rules, 
leading to the difference in $\rm TPR_\textit{avg}$). Notably, \ourtool{} with the self-hosted LLM (Qwen-Coder-2.5) also achieves a considerable $\rm TPR_\textit{avg}$ of 79.01\%, making it promising for privacy-sensitive and resource-constrained code-checking applications.

Based on the method rank, \ourtool{} significantly outperforms all baselines. Specifically, it achieves $2.93\times$ the performance of NoCaseLLM, $3.34\times$ of AllCasesLLM, $2.57\times$ of $\mathrm{NoCaseLLM}^R$, $2.80\times$ of $\mathrm{NoCaseLLM}^C$ and $2.11\times$ of $\mathrm{NoCaseLLM}^{RC}$ on $\rm TPR_\textit{avg}$. Though the performance of NoCaseLLM can be augmented with prompt engineering techniques (COT and RAG), the metric $\rm TPR_\textit{avg}$ is still below 40\%, and most generated checkers cannot even pass compilation. Compilation errors primarily stem from insufficient API knowledge, leading to API hallucinations such as incorrect class names and method calls. These results also prove that simply retrieving API-contexts based on the rule description (in $\mathrm{AutoChecker}^{R}$ and $\mathrm{AutoChecker}^{RC}$) is coarse-grained, often resulting in retrieval failures and, eventually LLM hallucinations.

\begin{table}[t]
\caption{Overall Performance Results of \ourtool{} and Baselines Using Different LLMs on the Benchmark RuleSet.}
\label{tab:RQ1_result}
\centering
\footnotesize
\addtolength{\tabcolsep}{-1pt}
\begin{tabular}{lccccc}
    \toprule
    \rule{0pt}{2.5ex}
    \multirow{2}{*}{\textbf{Method + LLM}} & \textbf{\#}$\textbf{Rule}_{pc}$ &
    \textbf{\#}$\textbf{Rule}_{pot}$ &
    \textbf{\#}$\textbf{Rule}_{pat}$ &
    \textbf{\#}$\textbf{TC}_{pass}$ & 
    \multirow{2}{*}{$\textbf{TPR}_{avg}$} \\
    
     & \textbf{(/20)} & \textbf{(/20)} & \textbf{(/20)} & \textbf{(/373)} &  \\ 
    
    \midrule
    \rowcolor{gray!20}
    \rule{0pt}{2.5ex}
    $NoCaseLLM$ & \multicolumn{5}{r}{\textit{\ding{46} naive baseline without test cases}}\\
    \rowcolor{gray!10}
    \ \ \ + Llama3.1  & 0 &0 & 0 & 0  & 0.00\% \\ 
    \rowcolor{gray!10}
    \ \ \ + Qwen2.5-Coder & 5   &5  & 1 & 40  & 19.41\% \\ 
    \rowcolor{gray!10}
    \ \ \ + GPT-4 & 7    &7  & 1 & 62  & 27.92\% \\
    \rowcolor{gray!10}
    \ \ \ + DeepSeek-V3\includegraphics[width=0.8em]{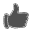} & 8   &8  &1  &  56  & 28.06\% \\
     
    \hdashline[2pt/2pt]
    
    \rowcolor{gray!20}
    \rule{0pt}{2.5ex}
    $AllCasesLLM$ & \multicolumn{5}{r}{\textit{\ding{46} naive baseline with all test cases}}\\
    \rowcolor{gray!10}
    \ \ \ + Llama3.1& 0  & 0  & 0 &  0  & 0.00\% \\ 
    \rowcolor{gray!10}
    \ \ \ + Qwen2.5-Coder& 4   & 4  & 1 & 17  & 14.40\% \\ 
    \rowcolor{gray!10}
   \ \ \ + GPT-4 & 5   &5  &2  &  36  & 21.53\% \\
   \rowcolor{gray!10}
   \ \ \ + DeepSeek-V3\includegraphics[width=0.8em]{figs/good.png}& 6  &6  &  2 &  43  & 24.60\% \\
     
    \Xhline{0.5px}
    \rowcolor{gray!30}
    \rule{0pt}{2.5ex}
    $NoCaseLLM^\textit{R}$ & \multicolumn{5}{r}{\textit{\ding{46} enhanced baseline with RAG}}\\
    \rowcolor{gray!20}
    \ \ \ + Llama3.1& 2   &2  &0   & 16  & 4.71\% \\ 
    \rowcolor{gray!20}
    \ \ \ + Qwen2.5-Coder & 9   & 9 & 2 &  60  & 30.68\% \\ 
    \rowcolor{gray!20}
   \ \ \ + GPT-4 & 10   &10  &  1 & 108  & 30.82\% \\
   \rowcolor{gray!20}
    \ \ \ + DeepSeek-V3\includegraphics[width=0.8em]{figs/good.png}& 9   & 9  & 2 & 92 & 32.05\% \\
     
    \hdashline[2pt/2pt]
    \rowcolor{gray!30}
    \rule{0pt}{2.5ex}
    $NoCasesLLM^\textit{C}$ & \multicolumn{5}{r}{\textit{\ding{46} enhanced baseline with COT}}\\
    \rowcolor{gray!20}
    \ \ \ + Llama3.1& 0   &0  & 0  & 0  & 0.00\% \\ 
    \rowcolor{gray!20}
    \ \ \ + Qwen2.5-Coder & 6   & 6  & 1 & 45  & 21.18\% \\ 
    \rowcolor{gray!20}
   \ \ \ + GPT-4 & 8   & 8  &1  & 94  & 27.26\% \\
    \rowcolor{gray!20}
    \ \ \ + DeepSeek-V3\includegraphics[width=0.8em]{figs/good.png} & 9   & 9  & 0 & 66  & 29.40\% \\
     
    \hdashline[2pt/2pt]
    \rowcolor{gray!30}
    \rule{0pt}{2.5ex}
    $NoCaseLLM^{\textit{RC}}$ & \multicolumn{5}{r}{\textit{\ding{46} enhanced baseline with RAG + COT}}\\
    \rowcolor{gray!20}
    \ \ \ + Llama3.1& 2   &  2  & 0 & 7  & 6.25\% \\ 
    \rowcolor{gray!20}
    \ \ \ + Qwen2.5-Coder & 9   & 9  & 1 & 60  & 30.49\% \\ 
    \rowcolor{gray!20}
    \ \ \ + GPT-4 & 9   & 9  &1  & 105  & 27.74\% \\
    \rowcolor{gray!20}
   \ \ \ + DeepSeek-V3\includegraphics[width=0.8em]{figs/good.png} & 11   & 11  & 1 & 101  & 38.93\% \\
     
    \Xhline{0.5px}
    \rowcolor{ManiBlue!20}
    \rule{0pt}{2.5ex}
    \textit{\footnotesize{$\textbf{AutoChecker}$}} & \multicolumn{5}{r}{\textit{\ding{46} our approach}}\\
    \rowcolor{ManiBlue!10}
    \textbf{\ \ \ + Llama3.1}& \textbf{3}  &\textbf{3} &\textbf{1} &  \textbf{22}  & \textbf{8.41\%} \\ 
    \rowcolor{ManiBlue!10}
    \textbf{\ \ \ + Qwen2.5-Coder} & \textbf{20} \textcolor{ManiPurple}{\ding{95}}  &\textbf{20} \textcolor{ManiPurple}{\ding{95}} & \textbf{4} & \textbf{257}  & \textbf{79.01\%} \\ 
    \rowcolor{ManiBlue!10}
    \textbf{\ \ \ + GPT-4}\includegraphics[width=0.8em]{figs/good.png} & \textbf{20} \textcolor{ManiPurple}{\ding{95}} & \textbf{20} \textcolor{ManiPurple}{\ding{95}} & \textbf{6} \textcolor{ManiPurple}{\ding{95}}&  \textbf{278} \textcolor{ManiPurple}{\ding{95}}  & \textbf{82.28\%} \textcolor{ManiPurple}{\ding{95}}\\
    \rowcolor{ManiBlue!10}
    \textbf{\ \ \ + DeepSeek-V3}& \textbf{19} & \textbf{19} & \textbf{4} & \textbf{278} \textcolor{ManiPurple}{\ding{95}}  & \textbf{80.86\%} \\
    \bottomrule
\end{tabular}
\footnotesize{We keep the result with higheset $\rm TPR_\textit{avg}$ across three runs for each method.\\ \#$\mathrm{TC}_{pass}$ denotes the number of passed test cases in total; \ding{95} marks the best result of each metric across all methods; \includegraphics[width=0.8em]{figs/good.png} is the best LLM (based on $\rm TPR_\textit{avg}$) for each method.}
\end{table}

\begin{figure}[!b]
\setlength{\abovecaptionskip}{5pt}
\setlength{\belowcaptionskip}{-5pt}
\centering
\includegraphics[width=0.46\textwidth]{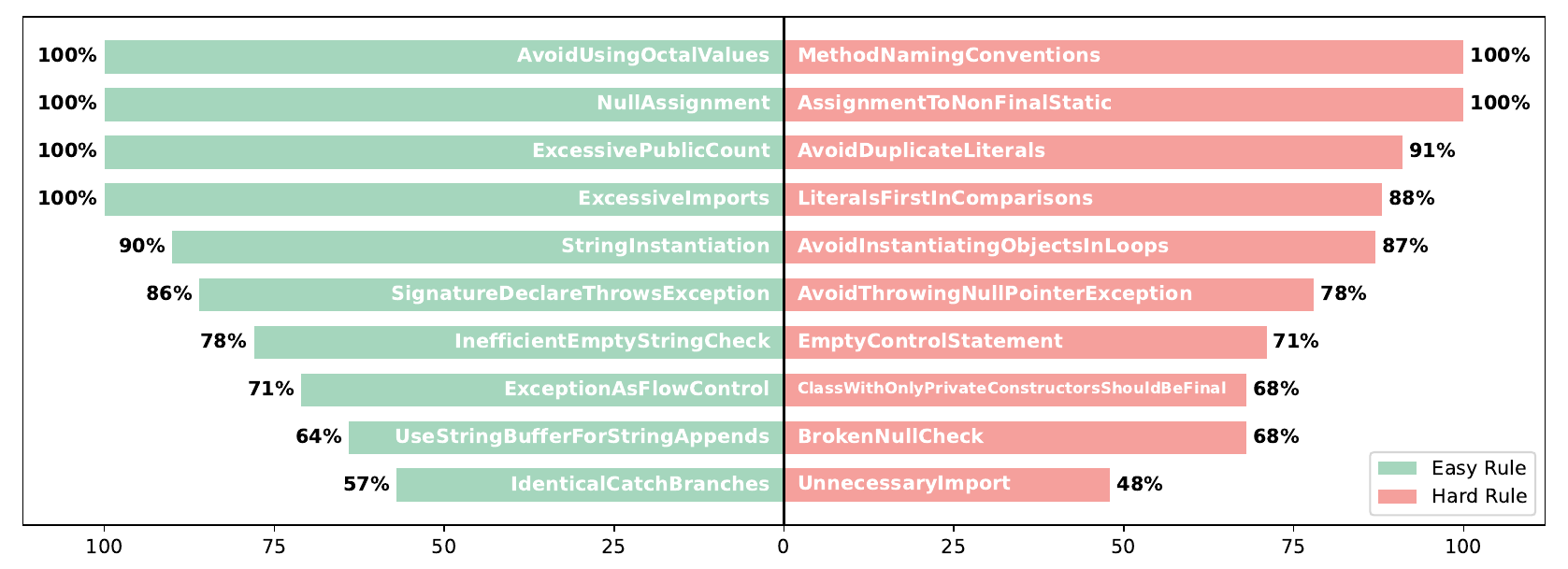}
\caption{TPR Distribution for Checkers Generated by \ourtool{}+GPT-4 on Easy Rules and Hard Rules}
\label{fig:TPR_distribution}
\end{figure}

To further analyze \ourtool{}'s performance on easy and hard rules, we collect the TPR distribution for all rules using GPT-4, the best-performing LLM. As shown in Fig.~\ref{fig:TPR_distribution}, the results align with expectations: hard rules are more challenging, with average TPRs of 84.60\% for easy rules and 79.90\% for hard rules. Specifically, the generated checkers pass all tests for 4 easy rules and 2 hard rules.

\paragraph{Failure Discussion}
From the results, checkers generated by \ourtool{} with GPT-4 fail on 95 test cases, which are skipped after reaching the retry limit. We randomly sample about half (45) from different rules and categorize the failures into compilation errors (due to hallucinated APIs), selected test failures (failing the current test), and regression test failures (failing previously passed tests). Besides API retrieval precision, LLM capability is also a key reason for these failures, as we observed LLMs using deprecated or wrong APIs even when correct ones are provided (e.g., using deprecated API \texttt{jjtGetNumChildren} for rule \textit{ExecptionAsFlowControlRule} even the correct API \texttt{getNumChildren} has been provided in prompts).

\begin{center}
\begin{tcolorbox}[colback=gray!10,
                  colframe=black,
                  width=0.48\textwidth,
                  arc=1mm, auto outer arc,
                  boxrule=0.5pt,
                  left=1pt, right=1pt, top=1pt, bottom=1pt
                 ]
\textbf{\ding{224} Answering RQ1:} \ourtool{} outperforms both naive and enhanced baselines, achieving the highest 82.28\% $\rm TPR_\textit{avg}$ with GPT-4. It indicates that our approach can effectively help developers to write their own checkers only with the rule and test suite.

\end{tcolorbox}
\end{center}

\subsection{RQ2: Ablation Study}

To evaluate the effectiveness of specific strategies in \ourtool{}, we conduct ablation experiments. As GPT-4 and DeepSeek-V3 achieve comparable performance (discussed in RQ1), we use both for the ablation study. Tab.~\ref{tab:RQ2_result} gives the overall results.

We start by analyzing the effectiveness of retrieval and iteration settings. In terms of $\rm TPR_{\textit{avg}}$, $\rm AutoChecker^{\textit{WoI}}$ achieves better performance using DeepSeek-V3, while $\rm AutoChecker^{\textit{WoR}}$ performs better using GPT-4. Compared to them, \ourtool{} with GPT-4 improves $\rm TPR_{\textit{avg}}$ by 53.97\% and 22.31\%, respectively. This shows that both API-context retrieval and the TDCD cycle are essential, with API-context retrieval being particularly crucial. As shown in the second column, $\rm AutoChecker^{\textit{WoI}}$ has fewer pass-compilation checkers than $\rm AutoChecker^{\textit{WoR}}$. Without accurate API knowledge, \ourtool{} and any other LLM-based methods use hallucinated APIs and will fail due to compilation errors.

To validate the effectiveness of the meta-settings (Meta-Op Set and Meta-API DB) in \ourtool{}, we introduce the ablation method $\rm AutoChecker^{\textit{WoI}}$. As shown in Tab.~\ref{tab:RQ2_result}, while it gets good performance on $\rm TPR_{\textit{avg}}$ of around 70\% only based on the Full-API DB, it is still at least 10 percent point lower than \ourtool{}. This result highlights the critical role of meta-settings in retrieval.

\begin{center}
\begin{tcolorbox}[colback=gray!10,
                  colframe=black,
                  width=0.48\textwidth,
                  arc=1mm, auto outer arc,
                  boxrule=0.5pt,
                  left=1pt, right=1pt, top=1pt, bottom=1pt
                 ]
\textbf{\ding{224} Answering RQ2:} Both the \textit{Retrieval} and \textit{Iteration} strategies are necessary for \ourtool{}. Also, with the meta-settings, its average test pass rate increases by around 10 percentage points.
\end{tcolorbox}
\end{center}

\begin{table}[t]
\caption{Results of \ourtool{} and Ablation Methods using GPT-4 and DeepSeek-V3 on the Benchmark Ruleset.}
\label{tab:RQ2_result}
\centering
\footnotesize
\addtolength{\tabcolsep}{-1pt}
\begin{tabular}{lccccc}
    \Xhline{1px}
    \rule{0pt}{2.5ex}
    \multirow{2}{*}{\textbf{Method + LLM}} & \textbf{\#}$\textbf{Rule}_{pc}$ &
    \textbf{\#}$\textbf{Rule}_{pot}$ &
    \textbf{\#}$\textbf{Rule}_{pat}$ &
    \textbf{\#}$\textbf{TC}_{pass}$ & 
    \multirow{2}{*}{$\textbf{TPR}_{avg}$} \\
    
     & \textbf{(/20)} & \textbf{(/20)} & \textbf{(/20)} & \textbf{(/373)} &  \\ 
     
    \Xhline{0.8px}
     \rowcolor{ManiRed!20}
     \rule{0pt}{2.5ex}
     $AutoChecker^{\textit{WoI}}$ & \multicolumn{5}{r}{\textit{\ding{46} ablation method without iterations}}\\
      \rowcolor{ManiRed!10}
    \ \ \ + GPT-4 & 8   & 8  & 2 & 65  & 29.37\% \\
    \rowcolor{ManiRed!10}
    \ \ \ + DeepSeek-V3\includegraphics[width=0.8em]{figs/good.png}& 14   & 14 &4  &  141  & 53.44\% \\

     \hdashline[2pt/2pt]
     \rowcolor{ManiRed!20}
     \rule{0pt}{2.5ex}
    $AutoChecker^{\textit{WoR}}$ & \multicolumn{5}{r}{\textit{\ding{46} ablation method without API-context retrieval}}\\
     \rowcolor{ManiRed!10}
    \ \ \ + GPT-4\includegraphics[width=0.8em]{figs/good.png} & 18   & 18  & 2 & 231  & 67.27\% \\
    \rowcolor{ManiRed!10}
    \ \ \ + DeepSeek-V3& 15   & 15 &2  &  221  & 59.17\% \\
    
    \hdashline[2pt/2pt]
     \rowcolor{ManiRed!20}
     \rule{0pt}{2.5ex}
    $AutoChecker^{\textit{WoM}}$ & \multicolumn{5}{r}{\textit{\ding{46} ablation method without Meta-Op Set and Meta-API DB}}\\
     \rowcolor{ManiRed!10}
    \ \ \ + GPT-4 & 17   & 17  & 3 & 256  & 66.42\% \\
    \rowcolor{ManiRed!10}
    \ \ \ + DeepSeek-V3\includegraphics[width=0.8em]{figs/good.png}& 18   & 18 & 1  & 258  & 72.92\% \\
    
    \Xhline{0.8px}
    \rowcolor{ManiBlue!20}
    \rule{0pt}{2.5ex}
    \textit{\footnotesize{$\textbf{AutoChecker}$}} & \multicolumn{5}{r}{\textit{\ding{46} our approach}}\\
    \rowcolor{ManiBlue!10}
    \textbf{\ \ \ + GPT-4}\includegraphics[width=0.8em]{figs/good.png} & \textbf{20} \textcolor{ManiPurple}{\ding{95}} & \textbf{20} \textcolor{ManiPurple}{\ding{95}} & \textbf{6} \textcolor{ManiPurple}{\ding{95}}&  \textbf{278} \textcolor{ManiPurple}{\ding{95}}  & \textbf{82.28\%} \textcolor{ManiPurple}{\ding{95}}\\
    \rowcolor{ManiBlue!10}
    \textbf{\ \ \ + DeepSeek-V3}& \textbf{19} & \textbf{19} & \textbf{4} & \textbf{278} \textcolor{ManiPurple}{\ding{95}}  & \textbf{80.86\%} \\
    \Xhline{0.8px}
\end{tabular}
\end{table}

\subsection{RQ3: Cost of \ourtool{}}
We evaluate \ourtool{}'s time and financial costs respectively. Our observations show consistent time and token costs across different LLMs, as they are all accessed via official or self-hosted APIs. Since \ourtool{} struggles to achieve good results with Llama3.1, we analyze the average costs across the other three LLMs.

\begin{figure}[!ht]
\centering
\setlength{\abovecaptionskip}{5pt}
\setlength{\belowcaptionskip}{-5pt}
\includegraphics[width=0.48\textwidth,]{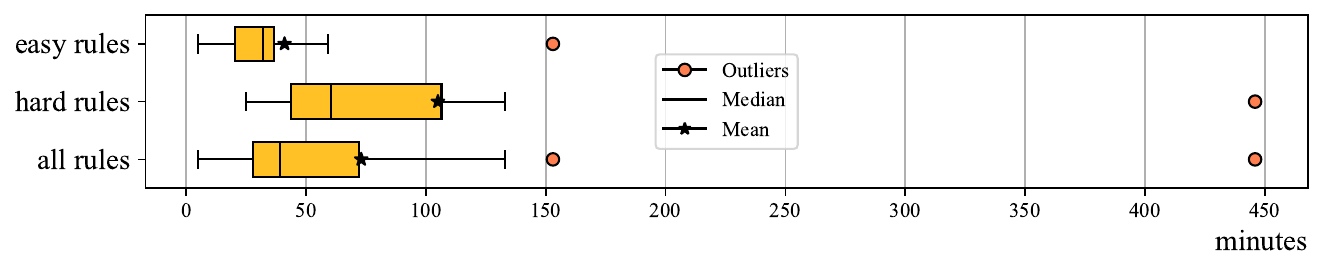}
\caption{Time Cost of AutoChecker on Different Rule Set}
\label{fig:Time Cost}
\end{figure}

For time cost, we measure the average duration across three runs for easy rules, hard rules, and all rules combined. As shown in Fig.~\ref{fig:Time Cost}, \ourtool{} takes 70 minutes to generate the final checker per rule on average: 40 minutes for easy rules and 100 minutes for hard rules. It is more efficient than traditional manual development, which often takes several days and involves multiple roles. 

For financial cost, we calculate the token usage (121k input and 388 output tokens on average) using the default tokenizers. Generating a checker costs approximately \$3.65 for GPT-4 and \$0.035 for DeepSeek-V3 per rule. As Tab.~\ref{tab:RQ1_result} shows, \ourtool{} achieves comparable performance across LLMs, enabling users to opt for cheaper options (DeepSeek-V3) or API-free ones (Qwen2.5-Coder). For enterprises that need custom checkers, the financial cost of \ourtool{} is far more affordable than manual developing.

\begin{center}
\begin{tcolorbox}[colback=gray!10,
                  colframe=black,
                  width=0.48\textwidth,
                  arc=1mm, auto outer arc,
                  boxrule=0.5pt,
                  left=1pt, right=1pt, top=1pt, bottom=1pt
                 ]
\textbf{\ding{224} Answering RQ3:}
The time and financial cost of \ourtool{} is more affordable compared to traditional checker development.


\end{tcolorbox}
\end{center}

\subsection{RQ4: Practicality in Real-world Projects}

To evaluate the applicability of \ourtool{}-generated checkers, we apply them to scan real-world projects and compare their performance with official checkers. For project selection, we select five popular Java projects\footnote{Algorithms/Java~\cite{algorithm}, elastic/elasticsearch~\cite{elasticsearch}, macrozheng/mall~\cite{mall}, google/guava~\cite{guava}, and spring-projects/spring-boot~\cite{springboot}.} from GitHub, each with over 50K stars and ranging from 50 to 1,517 KLOC. For checker selection, we use those that pass all tests and identify them as successful. Instead of selecting from a single run, we collect the successful checkers from all three runs. Specifically, we select the eight\footnote{The number of successful checkers across three runs: 6, 6, 5; Deduplicated total: 8.} successful checkers generated by \ourtool{} with GPT-4, as this exceeds the number with DeepSeek-V3 (six) and other LLMs.

\begin{table}[!ht]
\setlength{\abovecaptionskip}{5pt}
\caption{Violations Reported by the Official and \ourtool{}-Generated Checker on Real-world Projects}
\label{tab:practice}
\footnotesize
\setlength{\tabcolsep}{4pt}
\renewcommand{\arraystretch}{1.1}
\begin{tabular}{lc|r|r|r}
    \Xhline{0.8pt}
    \multirow{3}{*}{\textbf{Checker Rule}} & 
    \multirow{3}{*}{\textbf{\#TC$_{+}$}} &
    \multicolumn{3}{c}{\textbf{\#Violations on Five Projects}}
    \\
    \cline{3-5}
    & & 
    \textbf{official}  & 
    \multicolumn{2}{c}{\textbf{Checker$_\textit{AutoChecker+GPT-4}$}}
    \\
    \cline{4-5}
    & &  \textbf{checker} & \textbf{with TS$_\textit{orig}$} & \textbf{with TS$_\textit{aug}$} \\
    
    \Xhline{0.8pt}
    NullAssignment  & \textbf{+5}  &  2,560 & 1,632\ \textcolor{red}{($\downarrow$928)\textcolor{crashRed}{\ding{56}}}  & \textbf{2,562\ \textcolor{red}{($\uparrow$2)}} \\
    ExcessivePublicCount  & \textbf{+6}  &  389 & 330\ \textcolor{red}{($\downarrow$59)}  & \textbf{389\ \textcolor{dkgreen}{(=0)}} \\
    ExcessiveImports  &\textbf{+0}  &  3,321 & 3,321\ \textcolor{dkgreen}{(=0)} &  \textbf{3,321\ \textcolor{dkgreen}{(=0)}} \\
    AvoidUsingOctalValues  & \textbf{+7}  &  58 & 0\ \textcolor{red}{($\downarrow$58)}  & \textbf{58\ \textcolor{dkgreen}{(=0)}}  \\
    MethodNamingConventions & \textbf{+1}  &  11,562 & 11,560\ \textcolor{red}{($\downarrow$2)} & \textbf{11,562\ \textcolor{dkgreen}{(=0)}}  \\
    AssignmentToNonFinalStatic &\textbf{+0}  &  8 & 8\ \textcolor{dkgreen}{(=0)\textcolor{crashRed}{\ding{56}}} & \textbf{8\ \textcolor{dkgreen}{(=0)}}  \\
    StringInstantiation &\textbf{+0}  &  347 & 347\ \textcolor{dkgreen}{(=0)} & \textbf{347\ \textcolor{dkgreen}{(=0)}} \\
    InefficientEmptyStringCheck &\textbf{+2}  &  16 & 28\ \textcolor{red}{($\uparrow$12)} & \textbf{16\ \textcolor{dkgreen}{(=0)}}  \\
    \Xhline{0.8pt}
\end{tabular}
\footnotesize{\textbf{TS$_{orig}$}(original test suite) + \textbf{TC$_{+}$}(new test cases) $\rightarrow$ \textbf{TS$_{aug}$}(augmented test suite);\\
\textbf{Checker with TS$_\textit{orig}$/TS$_\textit{aug}$:} generated checker based on a rule and its TS$_\textit{orig}$/TS$_\textit{aug}$;\\
\textcolor{crashRed}{\ding{56}} denotes that the checker meets crash during project scan.}
\end{table}

Table~\ref{tab:practice} shows the number of reported violations by official and \ourtool{}-generated checkers for each project. As shown in the fourth column, only three of the generated checkers based on the original test suite achieve the same performance compared to official ones. Among all the eight checkers, we observe missing reports (FNs) for four checkers and mistaken reports for one checker (FP), while two checkers encounter crashes during code scanning. 

Through careful manual analysis, we identified two main reasons for the performance gap:  implementation bugs (crash) and omitted checking logic for corner cases (FPs and FNs). Implementation bugs are mostly simple, e.g., missing null checks or failing to perform type checking before casting. They are quickly fixed by directly asking LLMs to repair with bug reports. For the other type, FPs and FNs can be reduced by augmenting the original test suite. To address this, we craft test cases to cover missing checking scenarios.
The number of added cases is shown in the second column in Table~\ref{tab:practice}.

After bug fixes and test augmentation, the newly generated checkers successfully report all violations, matching the performance of official ones. Additionally, the \textit{NullAssignment} chekcer reports two more violations, which are repeated ones at the same location (other reports are not repeated). As they are redundant true violations, we do not take them as FPs.

\begin{center}
\begin{tcolorbox}[colback=gray!10,
                  colframe=black,
                  width=0.48\textwidth,
                  arc=1mm, auto outer arc,
                  boxrule=0.5pt,
                  left=1pt, right=1pt, top=1pt, bottom=1pt
                 ]
\textbf{\ding{224} Answering RQ4:} 
Given an adequate test suite, \ourtool{} can generate checkers with real-world performance comparable to official ones. \ourtool{} shifts the development effort from the challenging task of writing checkers to the more manageable task of designing test suites.
\end{tcolorbox}
\end{center}

%% file: tex/5_threats.tex
\section{Threats to Validity}
\label{sec:threats}

The primary threat is the scalability of \ourtool{}. Since \ourtool{} is implemented for PMD and Java code checking, it may not easily apply to other code-checking tools and programming languages. To address this, we design \ourtool{} with framework- and language-independent strategies. Specifically, we propose a general checker development cycle based on LLMs, extendable to other tools and languages, and introduce a Meta-Op Set for fine-grained API-context retrieval, sharable across frameworks and languages. Ideally, \ourtool{} can be adapted to any tool that supports custom AST-based checkers and all languages. During migration, the main human effort involves collecting available APIs and constructing API-context DBs. While API collection is unavoidable and hard to automate, we introduce the semi-automated DB-construction process in Section~\ref{sec:retrieval} to minimize manual effort.
 
Another threat is that the selected rules in the benchmark ruleset may not be representative. To mitigate this, we choose rules from PMD's built-in set, which are widely recognized as references. After classifying these rules by difficulty and targets, we randomly select rules to ensure balanced representation across both difficulty levels and categories, as introduced in Section~\ref{sec:bench}.
 

%% file: tex/6_related_work.tex
\section{Related Work}

\subsection{Code Checker Development }

Traditional studies for automated static analysis primarily focused on manually implementing checkers based on discovered bug patterns~\cite{chen2017characterizing, bian2018nar, zhang2023detecting}. For instance, Chen et al.~\cite{chen2017characterizing} summarized anti-patterns in logging code, and Zhang et al.~\cite{zhang2023detecting} designed bug patterns for exception handling. These patterns are then manually encoded as a static checker. While effective, manual checker implementation is time-consuming and requires significant expertise.

Recently, the advent of Machine Learning (ML) and LLMs has inspired researchers to analyze and scan code in automated or semi-automated ways~\cite{harer2018automated, austin2021program, zhou2024large}. Most studies directly apply ML models to detect various vulnerabilities, such as GNN-based Devign~\cite{zhou2019devign}, Transformer-based LineVul~\cite{fu2022linevul}, LLM-based Llm4Vuln~\cite{sun2024llm4vuln}, etc. However, these approaches mostly focus on function-level detection and only identify limited types of vulnerabilities, which are not effective at detecting vulnerabilities in real-world code~\cite{steenhoek2024comprehensive, ding2024vulnerability}.

In order to scan real-world projects, several approaches have been recently proposed to combine static analysis tools with LLMs. Specifically, Wang et al.~\cite{wang2023boosting} and Li et al.~\cite{li2024llm} leverage LLMs to infer source-sink specifications to augment taint checkers for a given project and CWE, while some studies~\cite{li2023assisting,chen2024utilizing,li2024enhancing} directly use LLMs to reduce the false positive alarms of static checking tools. However, these efforts focus on improving existing checkers rather than creating new ones. In contrast, \ourtool{}  generates custom checkers through an automated end-to-end way based on LLMs.

\subsection{LLM-based Repo-level Code Generation}

Recently, code-related tasks like code generation have been revolutionized by LLMs~\cite{fan2023large, hou2024large, kou2024large}. LLMs have shown incredible capability in generating programs~\cite{chen2021evaluating, liu2023your}. Repo-level code generation aims at generating code using the APIs defined in the repository~\cite{Zhang2023repocoder2023}. Compared to function-level generation, repo-level code generation is more challenging and downstream, requiring repo-specific API knowledge. A recent survey~\cite{deng2024r2c2} categorized methods for repo-level generation into two types:  fusion-based and ranking-based. 

Fusion-based approaches~\cite{shrivastava2023repofusion, Agrawal2023Monitor, ding2022cocomic} jointly model repo-context into the LLM. Among these studies, MGD~\cite{Agrawal2023Monitor} queried static analysis tools in the background, and the answers participated in the model's decoding stage to influence code generation. These approaches usually need to modify the model decoding process, while \ourtool{} augments context directly into the prompt.

Ranking-based methods~\cite{Zhang2023repocoder2023, liu2023codegen4libs, shrivastava2023repository, zhang2023retrieve, zan2022language} retrieve the most similar code context from the repository into the prompt, which are primarily used in most studies. For example, Liu et al.~\cite{liu2023codegen4libs} find relevant import statements and similar code snippets into the prompt for repo-level code generation, while Zhang et al.~\cite{Zhang2023repocoder2023} apply two-stage retrieval for fine-grained API retrieval. In \ourtool{}, the logic-guided API-context retrieval method is also ranking-based, with optimizing settings (the decomposed logic-guided retrieval and Meta-Op DB) specifically designed for checker generation.

%% file: tex/7_conclusion.tex
\section{Conclusions}
We propose \ourtool{}, an LLM-powered approach to automatically write static code checkers with the rule description and the corresponding test suite. To the best of our knowledge, this is the first attempt to explore test-guided static checker generation using LLMs. \ourtool{} employs a novel test-driven checker development process to incrementally generate and refine the checker case by case. During each round, it retrieves related API-contexts as additional knowledge for the LLM through the logic-guided API-context retrieval method. Experimental results show that \ourtool{}’s effectiveness outperforms baseline approaches across all the metrics, including the average test pass rate. Furthermore, with adequate test cases, \ourtool{} is able to generate checkers that perform nearly as well as official ground truth checkers in real-world projects.